\documentclass[aps,prl,twocolumn]{revtex4}

\usepackage{graphicx}
\usepackage{amsmath,amssymb,amsfonts}

\begin{document}

\title{Cosmological Black Hole Formation due to QCD and Electroweak Phase Transitions}

\author{J. I. Kapusta}
\author{T. Springer}
\affiliation{School of Physics and Astronomy, University of Minnesota,
Minneapolis, MN 55455}

\date{\today}

\begin{abstract}
We solve dynamical equations of motion to determine the conditions under which an over-dense region in the early universe will lead to collapse to a black hole, starting from horizon crossing of the over-dense region to the point of gravitational instability.  Here we focus on the sensitivity to QCD and electroweak phase transitions.  We then solve rate equations to determine the mass distribution of black holes in the present universe.  A second order phase transition or rapid crossover would have significant consequences only if the index of primordial density fluctuations $n > 1.25$.  However, a first order transition would lead to a black hole dominated universe for any realistic value of $n$ including $n=1$.   
\end{abstract}

\maketitle

Various symmetries of particle physics that are broken in the vacuum are restored at sufficiently high temperatures \cite{ourbook}.  These include restoration of chiral symmetry and deconfinement in the QCD sector at temperatures above about 200 MeV and restoration of the spontaneously broken electroweak symmetry at temperatures above about 100 GeV.  The impact of cosmological phase transitions have generally been found to be inconsequential.  For example, a first order QCD phase transition could in principle have affected nucleosynthesis \cite{nucleo}, but any reasonable estimate or calculation of the length scales for inhomogeneities in isospin or baryon density yield values \cite{CK92, ourbook} that are several orders of magnitude too small to have observable consequences \cite{KK}.  Similarly, the baryon asymmetry of the universe could in principle have been generated during an electroweak phase transition \cite{EWbaryo}, but the consensus is that one must go beyond the standard model to generate the observed ratio of baryons to photons.  In this Letter we shall provide theoretical calculations which strongly suggest that if either the QCD or electroweak transitions were first order, the early universe would have been dominated by black holes, and the evolution of the universe would have been very different from what it is currently believed to have been.

Cosmological formation of black holes due to density fluctuations has a long history going back to Zel'dovich and Novikov \cite{Zeldovich} and Hawking \cite{Hawking}.  Carr and Hawking \cite{Carr} estimated that when the over-density $\delta \rho/\rho$ exceeded a critical value proportional to the square of the sound speed $v_s^2$ within a particle horizon it is susceptible to gravitational collapse to a black hole.  Jedamzik \cite{Jedamzik} subsequently argued that since during a first order QCD phase transition the sound speed vanishes this would be a very efficient epoch to form black holes.  The physical reason is that $v_s^2 = dP/d\rho$, where $P$ is pressure and $\rho$ is energy density, and during the finite time interval during which the mixed phase exists there is a change in $\rho$ with no change in $P$, hence no gradient force to prevent collapse.  Rather than solving the full equations of general relativity for time-evolving fluctuations in an expanding universe, which are very computationally intense, we shall follow an approach used by Cardall and Fuller \cite{CF}.  This approach allows us to survey a wide class of equations of state and initial conditions.  

We track the evolution of individual over-dense regions from horizon crossing to the point at which they stop expanding, referred to as the turnaround point.  We model the regions as spherical and homogeneous and so are characterized by only two quantities: the epoch at which they enter the horizon and their density contrast $\delta = \delta\rho/\rho$ at this time.  We specify the moment at which the regions enter the horizon by the average energy density $\bar{\rho}_h$ of the universe at this time.  (Quantities with an over-bar are the average quantities for the universe.  Quantities without the bar refer to those within the over-dense region.  A subscript $h$ denotes those quantities evaluated at the time of horizon crossing.)  Thus $\rho_h = (1+\delta)\bar{\rho}_h$.

The metric for the background is that of a flat FRW universe
\begin{equation}
ds^2 = -dt^2 + R^2(t)[dr^2 + r^2(d\theta^2 +
\sin^2\theta d\phi^2)] \, ,
\end{equation}
where $R(t)$ is the scale factor of the universe which obeys the
Friedmann equation
\begin{equation}
\label{Friedmann}
\left(\frac{dR}{dt}\right)^2 = \frac{8\pi G}{3}\bar{\rho}(t) R^2(t) \, . 
\end{equation}	
To track the evolution of the over-dense regions we
apply the metric of a closed FRW universe to the over-dense regions
\begin{equation}
ds^2 = -d\tau^2 + S^2(\tau)\left[\frac{dr^2}{1-\kappa
r^2} + r^2(d\theta^2 + \sin^2\theta d\phi^2)\right]
\end{equation}
where $\tau$ is the time coordinate and $S(\tau)$ is the analog of the scale factor for the inner region. The Friedmann equation for the region is
\begin{equation}
\left(\frac{dS}{d\tau}\right)^2 = \frac{8\pi G}{3}\rho (\tau) S^2(\tau) - \kappa
\end{equation}		 
We solve for $\kappa$ by matching the inner and outer metrics at horizon crossing \cite{CF}.
\begin{eqnarray}
\label{BC}
S_h &=& R_h\\
\left(\frac{dS}{d\tau}\right)_h &=& \left(\frac{dR}{dt}\right)_h
\end{eqnarray} 
The evolution of the over-dense region is given by
\begin{equation}
\left(\frac{dS}{d\tau}\right)^2 = \frac{8\pi G}{3}[\rho(\tau) S^2(\tau) - \bar{\rho}_h R_h^2 \delta] \, .
\end{equation}
The over-dense region will stop expanding when the quantity in parenthesis vanishes.  This happens when
\begin{equation}
\label{TurnaroundPoint}
\rho_* S_*^2 = \bar{\rho}_h R_h^2 \delta 
\end{equation}
where the * denotes the quantity evaluated at this moment.
  
This approach must be supplemented with the Jeans condition to determine whether the region will collapse to a black hole.  If
\begin{equation}
d_h \frac{S_*}{S_h} > \frac{\pi}{k_J} \, ,
\end{equation}
where $d_h$ is the particle horizon at crossing and $k_J$ is the Jeans wavenumber, the region will collapse.  The particle horizon at crossing is found in the standard way
\begin{equation}
d_h(t) = R(t)\int_0^t \, \frac{dt'}{R(t')} \, .
\end{equation}
The relativistic Jeans wavenumber is given by \cite{Weinberg}
\begin{equation}
k_J = \sqrt{\frac{4\pi G(1+3v_s^2)w}{v_s^2}} \, ,
\end{equation}					  
where $w = P + \rho = Ts$ is the enthalpy. (The ratio of baryon density 
$n_{\rm B}$ to entropy density $s$ is of order $10^{-9}$ and so the baryon chemical potential is neglected here.)  The critical over-density can now be determined numerically if the equation of state is given.

If the speed of sound is constant the equation of state is $P = v_s^2 \rho$. Then the particle horizon distance can be solved analytically with the result that the critical over-density necessary for collapse is
\begin{equation}
\delta_c = \frac{8\pi^2}{3}  \frac{v_s^2}{(1+v_s^2)(1+3v_s^2)^3} \, .
\end{equation}
For radiation $v_s^2 = 1/3$ yielding $\delta_c = \pi^2/12 \approx 0.822$.  This is roughly consistent with the numerical results of Niemeyer and Jedamzik \cite{NJ_Numerical} who obtained $\delta_c = 0.70 \pm 0.02$ depending on the exact shape of the initial density perturbation when it crossed the horizon.  Musco, Miller and Rezzolla \cite{Musco} and Green, Liddle, Malik and Sasaki  \cite{Green} find $\delta_c = 0.45 \pm 0.02$ because they only consider the growing mode of the perturbation.  For this reason the critical deltas obtained using the semi-analytical approach as implemented here should be divided by two.  Thus for a purely radiation dominated universe $\delta_c = \pi^2/24 \approx 0.411$, in satisfactory agreement with Musco {\it et al.} and Green {\it et al.}.  On the other hand this factor of two may just be viewed as a phenomenological factor that better normalizes the results of the present approach with the more sophisticated numerical calculations.  Fortunately our principle conclusions concerning first order phase transitions are not sensitive to this factor of two.

There are many ways to parameterize an equation of state.  Perhaps the most physically intuitive is to write the entropy density as
\begin{equation}
s(T) = \frac{4\pi^2}{90} T^3 N_{\rm eff}(T)
\end{equation}
so that $N_{\rm eff}(T)$ represents an effective number of massless bosonic degrees of freedom.  Using a physically motivated parameterization of 
$N_{\rm eff}$ as a function of temperature one can then construct the pressure and energy density from thermodynamic identities.  For a first or second order transition we use the following \cite{ourbook}.
\begin{equation}
N_{\rm eff}(T) = \left\{ 
\begin{array}{ll}
N_2 - A_2 \, \exp\left\{-\frac{T - T_c}{\Delta_2}\right\} &\mbox{$T > T_c$} \\
N_1 + A_1 \, \exp\left\{\frac{T - T_c}{\Delta_1}\right\}  &\mbox{$T < T_c$}
\end{array}
\right. 
\end{equation}	
For consistency $N_1 + A_1 \le N_2 - A_2$; equality implies a second order transition and inequality implies a first order transition.  The $N_1$ is the number of degrees of well below $T_c$ and $N_2$ is the number well above.  If $\Delta_i \rightarrow 0$ this represents a bag model type of equation of state.  Although we have performed calculations for various strengths of first order, second order, and rapid crossover transitions we focus here on first order.  For QCD we assume three flavors of massless quarks plus gluons at high temperature and three species of massless pions at low temperatures, plus leptons and photons in both phases.   We choose $T_c = 170$ MeV and report here on bag model and softened first order with $\Delta_i = 0.05 T_c$ and $A_i = 11.125$.  For the second order transition we choose $A_1 = 13$ and $A_2 = 33.375$.  For electroweak we use a bag model-like equation of state with all degrees of freedom above $T_c = 100$ GeV massless and a latent heat density varying from $0.5 T_c^4$ to $1.5 T_c^4$. 

Figure 1 shows the critical value of over-density for QCD as represented by a bag model equation of state.  (Any strong first order equation of state yields a curve that looks very similar.)  The remarkable feature is the tail.  The top side of the tail is determined by the condition that the over-dense region just enters the mixed phase when it stops expanding, and the bottom side by the condition that the mixed phase is just ending when the over-dense region stops expanding.  Any over-dense region with initial values within the tail will stop expanding in the mixed phase and therefore collapse according to the Jeans criterion.  (For the bag model there is even an analytic expression for the tail.)  A second order or rapid crossover transition also have some vestige of this tail but it does not extend to infinite energy density or approach the x-axis since, although the sound speed is small over some range of energy density, it never vanishes.

\begin{figure}[t]
\includegraphics[width=6cm,angle=90]{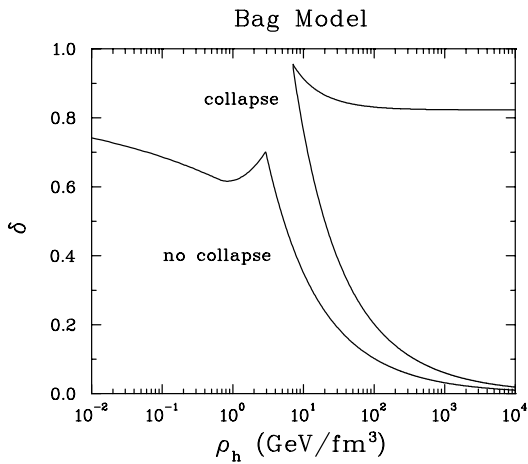}
\caption{\label{fig:1} Critical value of over-density as a function of energy density at horizon crossing for QCD.  Within the tail the over-dense region is in the mixed phase when it stops expanding ($dS/d\tau = 0$).  A 
radiation-dominated equation of state would have $\delta_c = 0.822$ independent of energy density.}
\end{figure}

The probability that a region which crosses the horizon (when the horizon mass is $M_h$) has an over-density $\delta$ is given by
\begin{equation}
\label{ProbDist}
P(\delta,M_h) = \frac{1}{\sqrt{2\pi}\sigma(M_h)}
\exp\left[-\frac{\delta^2}{2\sigma^2(M_h)}\right] \, . 
\end{equation}
The width $\sigma(M_h)$ is the COBE-normalized variance at the horizon mass $M_h$ taken to be \cite{GreenLiddle}
\begin{equation}
\sigma(M_h)=9.5\times 10^{-5} 
\left(\frac{M_h}{10^{22}M_{\rm sun}}\right)^{\frac{1-n}{4}} \, ,
\end{equation}
although it should be noted that Bringmann {\it et al.} \cite{Bringmann} contend that this is an overestimate.  Simple models of inflation predict the power spectrum for the perturbations to be a power law $\sim k^n$, where $k$ is the wavenumber associated with the perturbation and $n$ is referred to as the spectral index.

To find the present number density of black holes we must integrate over the history of the universe.  The number density at time $t$ is
\begin{eqnarray}
N(t) &=& \int_0^t dt' \left[ \frac{R(t')}{R(t)} \right]^3
\left| \frac{1}{V_h(t')} \frac{dV_h(t')}{dt'} \right|
\frac{1}{V_h(t')} \nonumber \\
&\times& \int_{\Delta(t')} d\delta \,
P(\delta,M_h(t')) \, .
\end{eqnarray}
The last factor is the probability that an over-density at horizon crossing will lead to collapse, where $\Delta(t')$ is the range of over-densities leading to collapse.  Working backwards, the next factor is the density factor (one per horizon crossing), then the horizon crossing rate, the dilution factor from formation to the observational time $t$, all of which are integrated from 0 to time $t$.  To compute the mass distribution $dN(M,t)/dM$ we insert the delta function $\delta \left( M_{\rm collapse}(\delta, M_h(t')) - M\right)$.  What this formula leaves out is the possibility that an existing black hole will find itself inside another collapsing over-dense region at some later time.  The observed spectrum of black holes is
\begin{equation}
\frac{dN^{\rm observed}(M,t)}{dM} = \frac{dN(M,t)}{dM} 
\exp\left( - B(M,t) \right)
\end{equation}
where
\begin{eqnarray}
B(M,t) &=& \int_0^t dt'
\left| \frac{1}{V_h(t')} \frac{dV_h(t')}{dt'} \right|
\int_{\Delta(t')} d\delta \,P(\delta,M_h(t')) \nonumber \\
&\times& \theta \left( M_{\rm collapse}(\delta, M_h(t')) - M\right) \, .
\end{eqnarray}
It is this observed distribution that we will subsequently refer to.

\begin{figure}[b]
\includegraphics[width=6cm,angle=90]{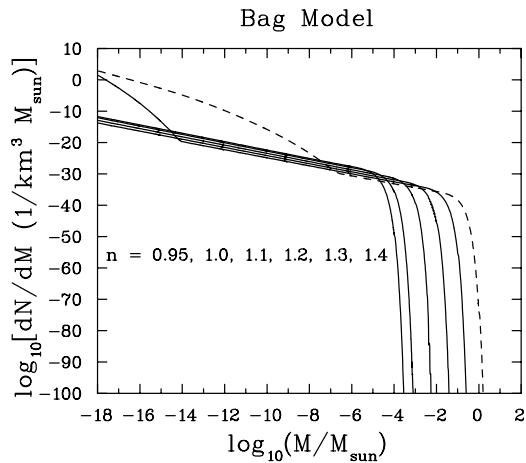}
\caption{\label{fig:2} Black hole mass spectrum at the present time due to a first order QCD phase transition for various values of spectral index $n$, starting with 1.4 (dashed line) in the upper left corner.}
\end{figure}

\begin{table*}[t*]
\noindent
Table I: Contribution of cosmologically produced black holes to the present value of $\Omega$.\\
\begin{tabular}{|c||c|c|c|c|c|c|c|}
\hline
\hspace*{10mm}& $n=1.4$ & $n=1.3$ & $n=1.25$ & $n=1.2$ & $n=1.1$ &
$n=1$ & $n=0.95$ \\
\hline
\hline
Bag Model & $7.8\times 10^{14}$ & $7.8\times 10^{11}$ & $4.4\times 10^7$ & $4.5\times 10^{7}$ & $4.6\times 10^{7}$ & $4.7\times 10^{7}$ & 
$4.7 \times 10^7$ \\
\hline
First Order & $1.3\times 10^{15}$ & $1.3\times 10^{12}$ & $2.3 \times 10^7$ &
$2.3\times 10^{7}$ & $2.3\times 10^{7}$ & $2.3\times 10^{7}$ & 
$2.2 \times 10^7$ \\
\hline
Second Order & $1.8\times 10^{15}$ & $1.9\times 10^{12}$
& $9.7\times 10^{-26}$ & $\sim 0$ & $\sim 0$ & $\sim 0$ & $\sim 0$ \\
\hline
Rapid Crossover & $1.8\times 10^{15}$ & $1.9\times 10^{12}$
& $9.7\times 10^{-26}$ & $\sim 0$ & $\sim 0$ & $\sim 0$ & $\sim 0$ \\
\hline
Fixed Speed & $1.8\times 10^{15}$ & $1.9\times 10^{12}$
& $9.7\times 10^{-26}$ & $\sim 0$ & $\sim 0$ & $\sim 0$ & $\sim 0$ \\
\hline
\end{tabular}
\end{table*}

Figure 2 shows the resulting black hole mass spectrum at the present time for QCD represented by a bag model equation of state.  The display begins at $10^{-18}$ solar masses since smaller black holes would have evaporated via Hawking radiation by now.  The sharp fall-off at large masses is related to the horizon mass during the epoch of the phase transition, although it does depend on the spectral index $n$.  Surprisingly, for $n$ in the range from 0.95 to 1.2 the spectrum is represented very well by the power-law $M^{-4/3}$.  (The deviation from this scaling law for larger $n$ is due to contributions from $\delta_c$ above the critical value for a fixed speed of sound.) All strong first order phase transitions have very similar features to these.

The contributions of cosmologically produced black holes to the present energy density of the universe, expressed in terms of canonical $\Omega$, are shown in Table I for a variety of equations of state.  We originally assumed a flat universe and implicitly assumed that black hole production would be a perturbation on the expansion of the universe.  Therefore any entry in the table with a contribution to $\Omega$ greater than 1.0 would indicate a universe dominated by black holes.  Since this is apparently not the case, a spectral index greater than 1.25 can be ruled out no matter which equation of state one looks at.  This is in close agreement with previous studies \cite{GreenLiddle}.  WMAP data suggest that $n$ is very close to 1.0, perhaps within a few percent \cite{WMAP}.  However, it must be noted that those observations probe comoving wavenumbers which are 10 to 20 orders of magnitude smaller than those of relevance here.  Even assuming values of $n$ close to 1 any strong first order QCD phase transition would have over-closed the universe by many orders of magnitude!  (Incidentally these results illustrate that the factor of two reduction in the $\delta_c$ hardly matters for the tails in Figure 1 since the width $\sigma(M_h)$ varies much more rapidly with small changes in $n$.)  We have reached the same conclusions with a first order electroweak phase transition.  For a latent heat density of $T_c^4$ the contribution to $\Omega$ at present is about $10^9$ for $0.95 \le n \le 1.20$, and scales with the latent heat.  The implication is that if either the QCD or the electroweak transition were of first order then the evolution of the early universe would have been dominated by black holes, contrary to prevailing knowledge.  How the universe would have evolved in such cases is beyond the scope of this paper.

Lattice gauge theory simulations of finite temperature QCD indicate that the physical strange quark mass is probably too large for QCD to undergo a first order phase transition, rather, it undergoes a very rapid crossover.  Analogous studies in the electroweak sector of the standard model indicate a first order phase transition only if the Higgs mass is less than around 70 GeV.  Since it is not, according to lack of observation of the Higgs particle at the Tevatron and at LEP, there is no first order phase transition in that sector either.  It is interesting to ponder the question of how the universe would have evolved had either the strange quark or the Higgs particle masses been smaller by a factor of 2 or 3.  More details concerning these calculations will be presented elsewhere.

We would like to thank Y.-Z. Qian, M. Peloso and L. L. R. Williams for discussions.  This work was supported by DOE grant DE-FG02-87ER40328.

\end{document}